# GaAs Microcavity Exciton-Polaritons in a Trap

Na Young Kim*[1,2], Chih-Wei Lai[1,3], Shoko Utsunomiya[3], Georgios Roumpos[1], Michael Fraser[1], Hui Deng[1], Tim Byrnes[3], Patrik Recher[1], Norio Kumada[4], Toshimasa Fujisawa[4], and Yoshihisa Yamamoto[1,3]

[1] *Edward L. Ginzton Laboratory, Stanford University, Stanford California, 94305 USA*

[2] *Institute of Industrial Science, University of Tokyo, Komaba Meguro-ku, Tokyo 153-8505, Japan*

[3] *National Institute of Informatics, Hitotsubashi Chiyoda-ku, Tokyo 101-8430, Japan*

[4] *NTT Basic Research Laboratories, NTT Corporation, Atsugi, Kanagawa 243-0198, Japan*

**Abstract.** We present a simple method to create an in-plane lateral potential in a semiconductor microcavity using a metal thin-film. Two types of potential are produced: a circular aperture and a one-dimensional (1D) periodic grating pattern. The amplitude of the potential induced by a 24 nm-6 nm Au/Ti film is on the order of a few hundreds of μeV measured at 6 ~ 8 K. Since the metal layer makes the electromagnetic fields to be close to zero at the metal-semiconductor interface, the photon mode is confined more inside of the cavity. As a consequence, the effective cavity length is reduced under the metal film, and the corresponding cavity resonance is blue-shifted. Our experimental results are in a good agreement with theoretical estimates. In addition, by applying a DC electric voltage to the metal film, we are able to modify the quantum well exciton mode due to the quantum confined Stark effect, inducing a ~ 1 meV potential at ~ 20 kV/cm. Our method produces a controllable in-plane spatial trap potential for lower exciton-polaritons (LPs), which can be a building block towards 1D arrays and 2D lattices of LP condensates.

## 1. Introduction

Coupling electromagnetic fields with massive particles and manipulating their coupling strength have been of great interest in solid-state systems [1]. The interests lie in the investigation of fundamental physical questions, for example coherence and decoherence properties of coupled systems, and in new designs of novel optoelectronic devices such as vertical-cavity surface emitting lasers. Diverse shapes of optical microcavity structures have been implemented in order to confine the electromagnetic fields [2]. A Fabry-Perot planar microcavity is a common choice to incorporate a quantum well structure. It is composed of a pair of distributed Bragg reflectors (DBRs) from alternating semiconductor layers of different refractive indices. In such a structure, when cavity photon fields and quantum well exciton modes are strongly coupled, new quasiparticles called microcavity exciton-polaritons emerge as a manifestation of a strong light-matter interaction [3].

Microcavity exciton-polartions have been regarded as composite bosons at the low density limit and a promising solid-state candidate to observe novel physical phenomena such as Bose-Einstein condensation (BEC) and superfluidity. Microcavity exciton-polaritons are advantageous for the study of the aforementioned phenomena at elevated temperatures because of their lighter mass arising from their photonic component. These particles have two distinct energy branches: upper exciton-polaritons and lower exciton-polaritons (LPs). Recently, evidence of BEC in LPs has been reported in terms of ground-state population, the thermal equilibrium to lattice, spontaneous long-range spatial and increased temporal coherence properties [4-8].

Along with the search for solid-state BEC systems, many efforts to produce further confinement potentials in planar systems have been attempted because additional confinement would help to determine the definite ground state energy and to facilitate the control of light-matter interactions. Influencing either the exciton mode energy or the cavity photon mode energy can create local in-plane potentials. Strain shifts the exciton energy, inducing a harmonic trap potential as an example of the former [9, 10]. The photonic component can be modified by modulating the cavity layer thickness in a mesa structure [11, 12]. Here we present a relatively simple way to form an in-plane potential by patterning a metal grating for a one-dimensional periodic potential and a circular aperture structure for a trap in two dimensions (2D). Under the metal film, the effective cavity length is reduced. Therefore, the energy of the cavity photon mode increases, while the exciton mode remains unchanged. In addition, applying an electric field perpendicular to the quantum wells causes a quantum confined Stark shift on the excitonic component, leading to the red-shift in the zero-momentum LP energy. This method has several advantages: (1) the in-plane potentials are created non-invasively after the full growth of a whole microcavity structure so that the non-radiative lifetime of the QW exitons is not increased; (2) the dimension of the in-plane potentials can be easily sub-microns by current advanced lithographic techniques; (3) various geometries in 1D and 2D of local in-plane potentials can be readily fabricated to explore many-body interaction effects among exciton-polaritons.

## 2. Experiments

Our λ/2 cavity contains three-stack of four GaAs quanum wells (QWs) located at the central antinodes of the cavity. The $Al_{0.15}Ga_{0.85}As/AlAs$ cavity is formed by a 16 -layer top DBR and a 20-layer bottom DBR. GaAs QWs (6.8 nm in width) are separated by 2.7 nm-AlAs barriers. We studied two samples with different substrates: the substrate of sample 1 is semi-insulating, whereas that of sample 2 is n-doped. A spatial inhomogeneity caused by a tapering in the layer thickness of the wafer allows us to tune the cavity resonance. The vacuum Rabi splitting energy is around 15 meV. On the surface of the top DBR, a grating or a hole is patterned by electron beam or photo-lithography, followed by a 24/6 nm Au/Ti metal deposition and a lift-off (Fig. 1 (a)). The grating pattern consists of repeated pair of a metal finger and a gap. Various sizes of the finger and the gap are used from 1.4 μm to 7.2 μm, and hole diameters vary from 10 μm to 90 μm.

Figure 1 (b) illustrates the schematic of our experimental setup. We measure time-integrated photoluminescence (PL) of LP emissions at 6 ~ 8 K as a response to a 3 ps pulsed excitation at ~ 60 degree by a Ti:sapphaire laser (~ 776 nm for sample 1 and ~ 768 nm for sample 2). The excitation spot is focused on the sample surface whose diameter is around 20 ~ 100 μm. The LP emission signals are collected in the normal direction by a high numerical aperture (NA = 0.55) objective lens. We use a standard μ-PL setup to access near-field (coordinate space) or far-field (momentum space) imaging and spectroscopy with a repositionable lens behind the objective lens. The in-plane potential strength is obtained from the LP emission energy while the laser excitation power is kept below the quantum degeneracy threshold.

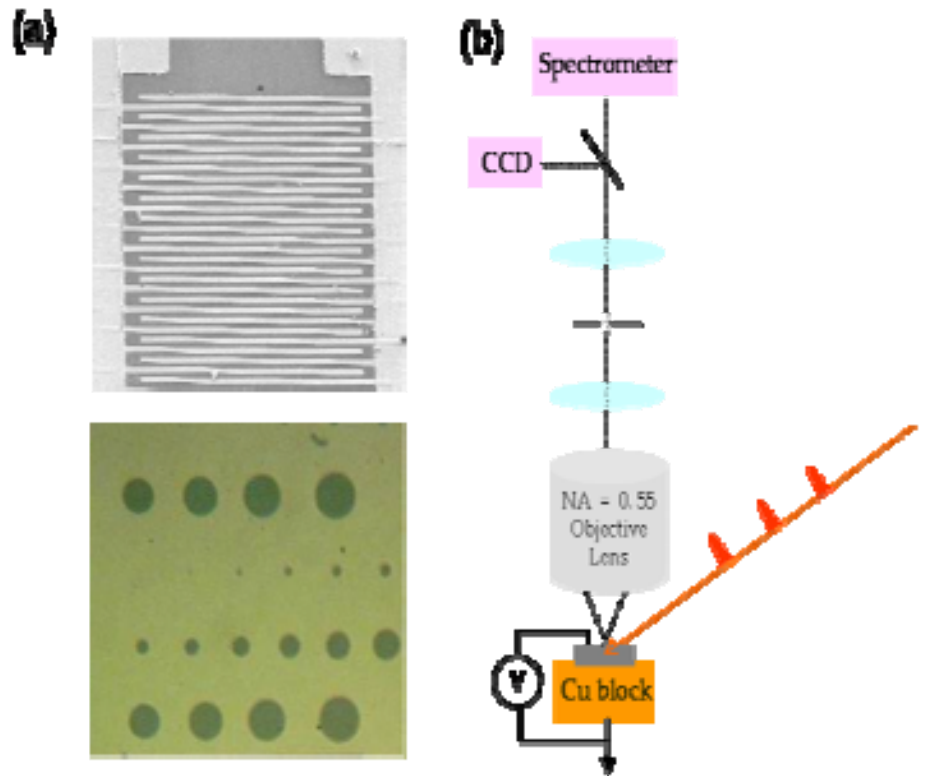

**Figure 1** (a) (Top) Scanning electron microscopy image of a one-dimensional grating pattern (pairs of a 1.4 μm metal finger and a 1.4 μm gap). (Bottom) Arrays of circular-aperture metal films. (b) A schematic of the near-field and the far-field imaging and spectroscopy setup.

### 3 Results and discussion

Representative near-field spectra of the periodic grating configuration and the hole structure are presented in Fig. 2. We excite the sample surface at low laser pump power below quantum degenerate threshold with a laser spot size of around 100 μm (Fig. 2 (a)) and 20 μm (Fig. 2 (b)). Figure 2 (a) and (b) are captured with a 5.6 μm metal-5.6 μm gap device and a 10 μm circular aperture, respectively. The PL intensity in Fig. 2 (a) shows a clear spatial modulation, and its peak-to-peak oscillation period is ~ 10.4 μm, which agrees well with the grating periodicity 11.2 μm to within a 7 % error. Since the LP emission at 780 nm through the metal film is masked and its PL transmittance is reduced, we measured the PL intensity through the metal reduced to 17 - 24 % of that above the bare surface. Thus, we lose ~ 41 – 49 % of the light emissions each time they pass the metal film. We extract the peak wavelength and the PL intensity of individual spectrum at each position, plotting them along position (Fig. 2 (c)-(d)). An anti-correlated trend between the PL intensity and the LP energy is clearly observed. The LP emission energy from the metal film is blue-shifted by 400 μeV on the average for this particular device. For the circular aperture, we measured the LP spectra at four symmetric spatial locations (marked with a cross symbol in Fig. 2 (d)) on the metal film away from the open aperture by ~ 50 – 100 μm. In this way, we can take into account of the spatial inhomogenous detuning paramter. The average of these four LP spectra above the metal film would provide the LP energy at the center of the circular aperture in the case the metal film is covered. Considering the reduced tranmittance through the metal film, we double the excitation laser power above the metal film, extracting the LP energy difference ($E_{Metal}$ - $E_{bare}$) between the metal film and the bare surface. $E_{Metal}$ is denoted as the LP energy above the metal film and $E_{bare}$ represents the LP energy measured from the bare surface.

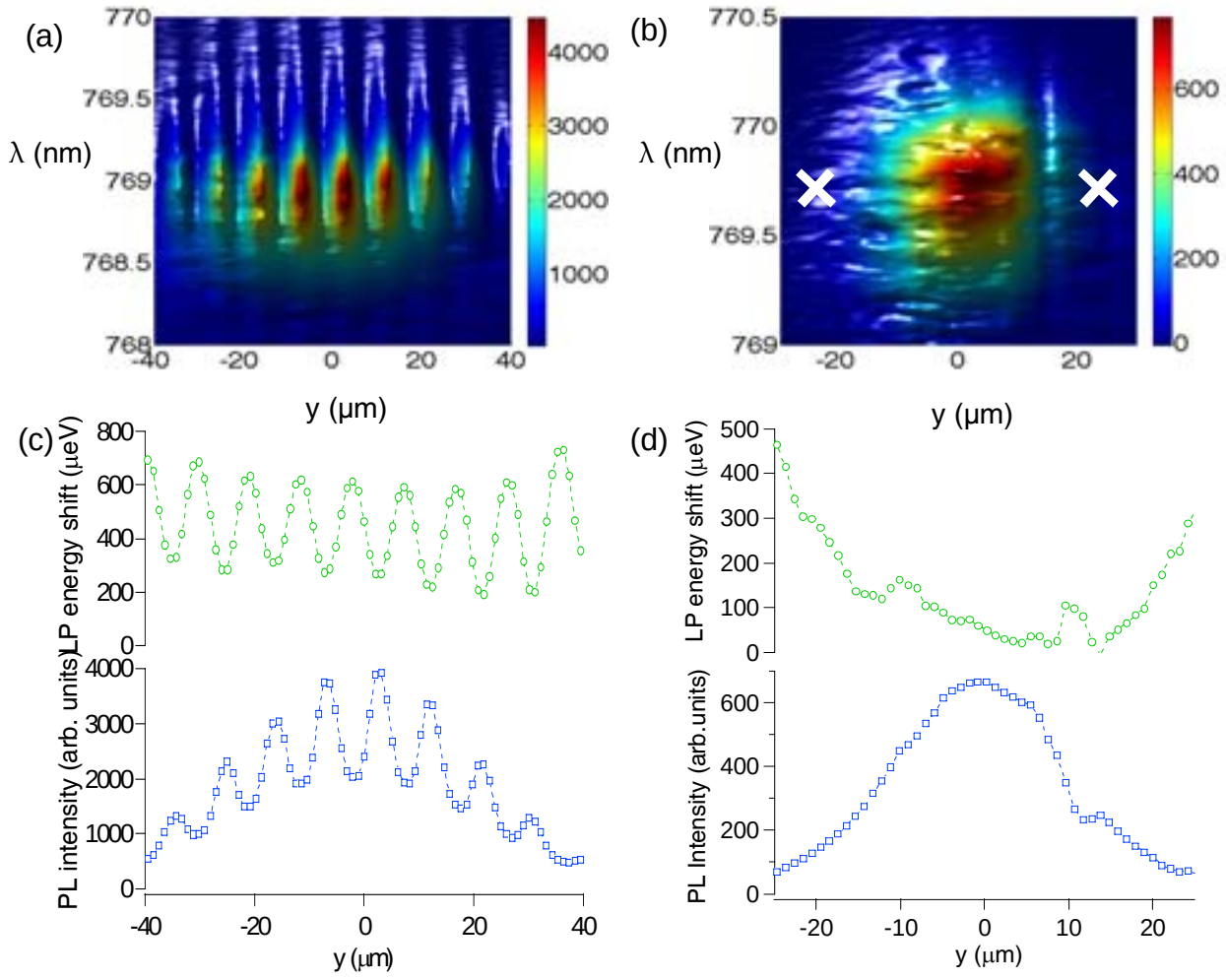

**Figure 2** Two-dimensional spatially-resolved near-field spectroscopy of (a) a 5.6 μm one-dimensional grid and (b) a 10 μm-diameter hole at P = 1 mW in sample 2. A color bar represents the lower-polariton (LP) photoluminescence (PL) intensity. The LP energy shift and its PL intensity in real space for (c) the 5.6 μm grid and (d) the 10 μm aperture at P = 1 mW. The LP energy shift represents the LP energy shifted by the minimum LP energy in space (1.6092 eV for (c) and 1.6077 eV for (d)). The cross marks in (d) indicate measurement positons above the metal film. Together with two more spatial points in x-axis, the average of the LP energy through metal films at four spatially symmetric positions provides the potential difference between the metal and the bare surface.

We compute the effect of a thin metal film on the LP energy and intensity using the transfer matrix method [13]. On the bare surface, the anti-node of the resonant cavity photon wavefunction sits at the air-semiconductor interface and the cavity photon field extends to the air. On the other hand, when the metal film is deposited on the DBR, the cavity photon field is attenuated through the metal layer, squeezing the cavity photon mode into the inner structure and pinning the cavity photon field at the metal-semiconductor interface to close to zero. Figure 3 illustrates the effect of the 30 nm metal film on the optical fields compared to the bare surface case. Consequently, the effective cavity length is reduced under the metal film and its resonant photon energy and corresponding LP energy increase (blue-shift). At zero detuning where the cavity photon energy is resonant with the QW exciton energy, the cavity photon mode energy is blue shifted by 400 μeV and the LP energy shift is expected to be roughly half size, 200 μeV since the photonic component of the LP at zero detuning is 50 %. Using the complex refractive indices of Au and Ti at 800 nm ($n_{Au}$ = 0.15 -4.86$i$, $n_{Ti}$ = 3.03-3.65$i$), the attenuation constants for Au and Ti are $\alpha_{Au}=2\pi \mathrm{Im}(n_{Au})/\lambda \sim 0.039$ nm$^{-1}$, $\alpha_{Ti}=2\pi \mathrm{Im}(n_{Ti})/\lambda \sim 0.029$ nm$^{-1}$. In addition, we consider reflection between interfaces. Overall the amplitude of emission through a 25 nm- Au metal film at 778 nm would be reduced to ~ 40 %, which corresponds well to the measured value.

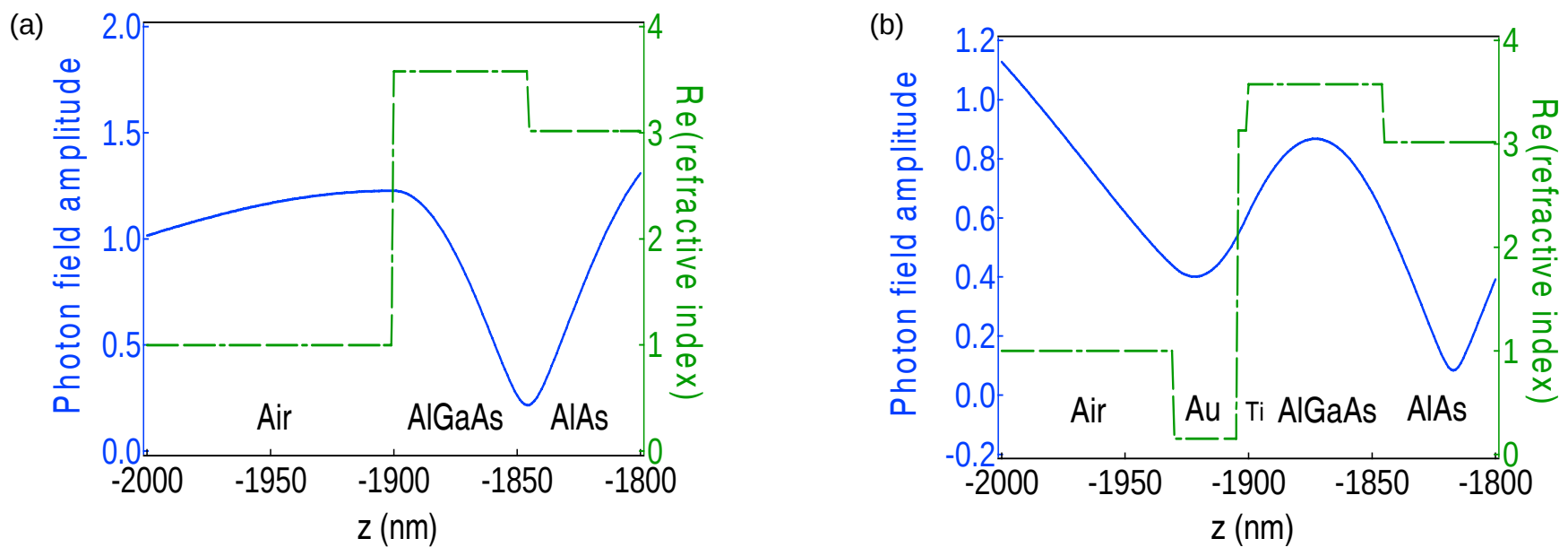

**Figure 3** Computed electromagnetic field amplitudes for a standing wave at resonance (blue) from the bare surface (a) and from the 26-4 nm Au/Ti metal layer (b) along the growth direction in z-axis. z is the distance from the center of the cavity. The real parts of refractive indices of materials are drawn in green. Only the region near the surface is shown here.

Theoretical model further predicts that the amount of the LP energy shift depends on the detuning ($\Delta$) at a given metal thickness. The more negative $\Delta$ is, namely, when the QW exciton energy ($E_{exc}$) is much larger than the cavity photon energy ($E_{cav}$), the more the LP energy shift is increased. Suppose the cavity photon energy shift to be $\delta$ induced by the metal film. The LP energy at $k_{//} = 0$ through the meal film is written as

$$E_{Metal} = \frac{1}{2}\left[E_{cav} + \delta + E_{exc} - \sqrt{(2g)^2 + (E_{cav} + \delta - E_{exc})^2}\right]$$

, where $2g$ is the vacuum Rabi splitting. Inserting $\Delta = E_{cav} - E_{exc}$, the energy difference between the metal and the bare surface is derived

$$E_{Metal} - E_{bare} = \frac{1}{2}\left[\delta - \sqrt{(2g)^2 + (\Delta + \delta)^2} + \sqrt{(2g)^2 + \Delta^2}\right]$$

The computed LP energy shift ($E_{Metal} - E_{bare}$) is plotted along the detuning in red (Fig. 4) with $g \sim 6.97$ meV, $E_{exc} \sim 1.59241$ eV and $\delta \sim 0.4$ meV. The values of $2g$ and $E_{exc}$ are obtained from the fitting results to the multiple dispersion curves obtained at various locations in sample 1 (not shown). Figure 4 summarizes the outcome of the the LP energy shift values from experimental data with sample 1 as well as the theoretical estimates.

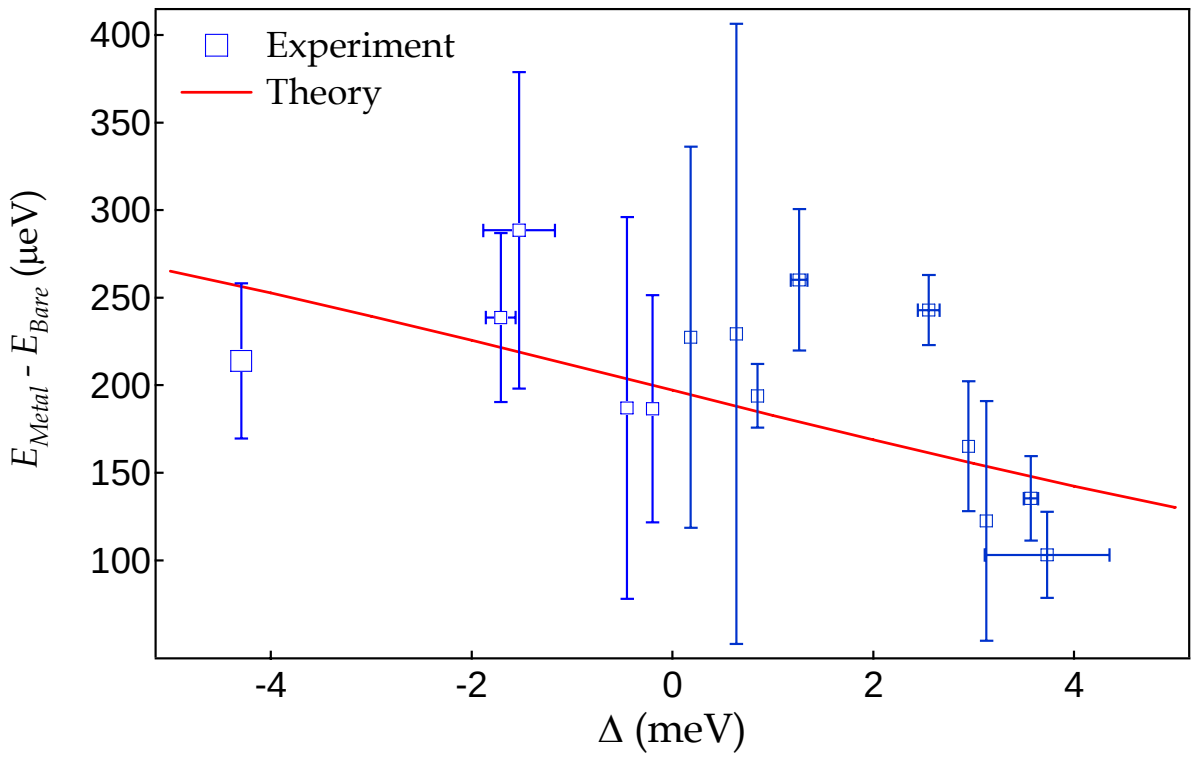

**Figure 4** The LP energy shift through the Au-Ti metal film deposited on the top surface of sample 1 ($E_{Metal}$) compared to the bare surface ($E_{bare}$) as a function of the detuning parameter (Δ). Sample 1 has a vacuum Rabi splitting energy 2g ~ 2×6.97 meV and $E_{exc}$ ~ 1.59241 eV.

The presence of the metal film modifies the cavity photon mode. In addition, we can modify the QW exciton mode by applying a DC voltage to the metal film across the wafer. The electric field perpendicular to the QW layer causes a quantum confined Stark effect [12]. Under the electric field, the energy difference between the ground state of the conduction band and that of the valence band is reduced. Therefore, the QW exciton energy is reduced (red shift). On the other hand, since an electron wavefunction in the conduction band and a hole wavefunction in the valence band are separated in space, the overlap of the electron and the hole wavefunctions is reduced, making the dipole moment of excitons smaller and yielding a reduced exciton binding energy. The coupling between the QW excitons and the cavity photons becomes weaker. The weaker coupling will cause the LP energy increased (blue shift). We compute the LP energy shift using a variational approach for exciton wavefunctions which takes into account the aforementioned electric field effect [15]. We assume that only heavy-hole excitons are relevant and we ignore light-hole excitons. Using the values of the GaAs Qw parameters at low temperature, we find that the overall LP energy becomes red shifted.

We apply the DC voltage to the metal film of sample 2 whose substrate is heavily n-doped. This maximum value (10 V) corresponds to ~ 20 kV/cm across the 4.7 μm insulating layer thickness. Experimentally, we observe that the trend of the LP energy shift is symmetric in both polarities of the DC electric field and the LP energy is systematically decreased by ~ 1 meV at the maximum applied electric field. For low electric fields, the measured LP energy shift can be explained by the theoretical estimate very well. However, we find that the LP energy in experiment is much more rapidly decreasing than that in theory. Investigation of this discrepancy will be left for future work.

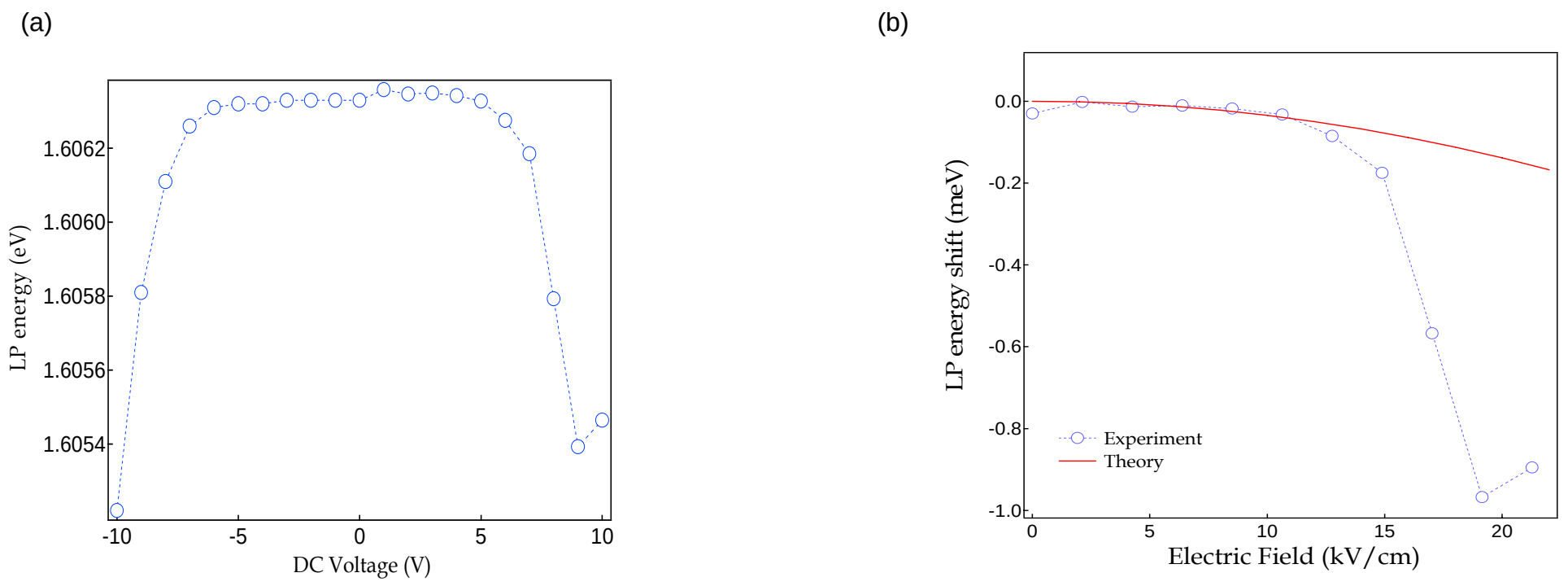

**Figure 5** (a) The LP energy shift due to the DC electric field across the wafer whose substrate is heavily n-doped. We apply the voltage up to 10 V across a 4.7 μm-wafer. (b) The experimental LP energy shift taken at $k_{//} = 0$ (circle) and the theoretical LP energy shift (red).

## 4. Conclusion

We show that a simple deposition of a metal thin-film on the top surface can modify both the cavity photon mode energy and QW exciton mode energy. In the former, the metal film layer acts as a mirror to define a reduced cavity length, causing the higher LP energy under the film. This can create an effective local in-plane potential in the range of around hundreds of μeV. We also show that the QW exciton mode energy can be controlled by the applied DC electric field perpendicular to the QW layer, yielding a ~ 1 meV LP energy shift at ~ 20 kV/cm. We envision that this method provides a simple but robust way to produce a controllable spatial potential to trap microcavity exciton-polaritons. We can extend readily this method to design and fabricate various patterns of geometries in 1D and 2D.

**Acknowledgements** This work is supported by JST/SORST, NTT and Special Coordination Funds for Promoting Science and Technology in University of Tokyo.

## References

[1] S. Haroche and D. Kleppner, Physics Today **42**, 24 (1989)

[2] K. J. Vahala, Nature **424**, 839 (2003).

[3] C. Weisbuch, M. Nishioka, A. Ishikawa, and Y. Arakawa, Phys. Rev. Lett. **69**, 3314 (1992).

[4] H. Deng, G. Weihs, C. Santori, J. Bloch, and Y. Yamamoto, Science **298**, 199 (2002).

[5] H. Deng, G. Weihs, D. Snoke, J. Bloch, and Y. Yamamoto, Proc. Nat. Acad. Sci. **100**, 15318 (2003).

[6] H. Deng, D. Press, S. Götzinger *et al*. Phys. Rev. Lett. **97**, 409 (2006).

[7] J. Kasprzak, M. Richard, S. Kundermann, A. Baas, P. Jeambrun, J. M. J. Keeling, F. M. Marchetti, M. H. Szymańska, R. André, J. L. Staehli, V. Savona, P. B. Littlewood, B. Deveaud, and Le Si Dang, Nature **443**, 409 (2006).

[8] H. Deng, G. S. Solomon, R. Hey, K. H. Ploog, and Y. Yamamoto, Phys. Rev. Lett. **99**,

126403 (2007).

[9] R. B. Balili, D. W. Snoke, L. Pfeiffer, and K. West, Appl. Phys. Lett. **88**, 031110 (2006).

[10] R. B. Balili, D. W. Snoke, L. Pfeiffer, and K. West, Science **316**, 1007 (2007).

[11] O. E. Daïf, A. Baas, T. Buillet, J. –P. Brantut, R. I. Kaitouni, J. L. Staehli, F. Morier-Genoud, and B. Deveaud, Appl. Phys. Lett. **88**, 061105 (2006).

[12] R. Idrissi Kaitouni *et al.* Phys. Rev. B **74**, 155311 (2006).

[13] P. Yeh, Optical waves in layered media. (John Wiley, Hoboken, N. J., 2005)

[14] D. A. B. Miller, D. S. Chemla, T. C. Damen, A. C. Gossard, W. Wiegmann, T. H. Wood, and C. A. Burrus, Phys. Rev. Lett. **53**, 2173 (1984).

[15] P. Harrison, Quantum Wells, Wires and Dots. (John Wiley, Hoboken, N. J., 2000)